\documentclass[conference]{IEEEtran}
\IEEEoverridecommandlockouts
\usepackage{cite}
\usepackage{amsmath,graphicx}
\usepackage{amssymb,amsfonts}
\usepackage{graphicx}
\usepackage{textcomp}
\usepackage{xcolor}
\usepackage{subfigure}
\usepackage{textcomp} 
\usepackage{siunitx}  
\usepackage{algorithm}
\usepackage{algpseudocode}
\usepackage{xpatch}
\usepackage{pgfplots}
\usepackage{pgfplotstable}
\usepackage{readarray}
\usepackage{siunitx}
\usepackage{bm}
\usepackage[nolist]{acronym}
\usepackage{bbm}
\usepackage{url}

\definecolor{TUMBeamerYellow}    {rgb} {1.000,0.706,0.000}    
\definecolor{TUMBeamerOrange}    {rgb} {1.000,0.502,0.000}    
\definecolor{TUMBeamerRed}       {rgb} {0.898,0.204,0.094}    
\definecolor{TUMBeamerDarkRed}   {rgb} {0.792,0.129,0.247}    
\definecolor{TUMBeamerBlue}      {rgb} {0.000,0.600,1.000}    
\definecolor{TUMBeamerLightBlue} {rgb} {0.255,0.745,1.000}    
\definecolor{TUMBeamerGreen}     {rgb} {0.569,0.675,0.420}    
\definecolor{TUMBeamerLightGreen}{rgb} {0.710,0.792,0.510}    

\DeclareMathOperator{\tr}{tr}
\DeclareMathOperator{\T}{T}
\DeclareMathOperator{\He}{H}

\DeclareMathOperator{\inv}{-1}

\DeclareMathOperator{\diag}{diag}

\newcommand{\eye}{\bm{\mathrm{I}}}
\newcommand{\bbthet}{\bm{\bar{\theta}}}
\newcommand{\bD}{\bm{D}}
\newcommand{\Her}{{\He}}
\newcommand{\td}{{\text{d}}}

\newcommand{\tdk}{{\text{d},k}}

\newcommand{\bC}{\bm{C}}
\newcommand{\bQ}{\bm{Q}}

\newcommand{\NB}{N_{\text{B}}}

\newcommand{\NRe}{N_{\text{R}}}

\newcommand{\tre}{{\text{r}}}

\newcommand{\trek}{{\text{r},k}}
\newcommand{\ts}{{\text{s}}}
\newcommand{\bH}{\bm{H}}

\newcommand{\bh}{\bm{h}}
\newcommand{\mThet}{\bm{\Theta}}

\newcommand{\bthet}{\bm{\theta}}

\newcommand{\cmplx}[1]{\mathbb{C}^{#1}}
\newcommand{\norm}[1]{\|#1\|}
\newcommand{\abs}[1]{|#1|}

\newcommand{\summe}[2]{\sum_{#1}^{#2}}

\def\BibTeX{{\rm B\kern-.05em{\sc i\kern-.025em b}\kern-.08em
    T\kern-.1667em\lower.7ex\hbox{E}\kern-.125emX}}
\begin{document}

\begin{acronym}
    \acro{AoD}{angle of departure}
    \acro{AoA}{angle of arrival}
    \acro{ULA}{uniform linear array}
    \acro{CSI}{channel state information}
    \acro{LOS}{line-of-sight}
    \acro{EVD}{eigenvalue decomposition}
    \acro{BS}{base station}
    \acro{MS}{mobile station}
    \acro{mmWave}{millimeter wave}
    \acro{DPC}{dirty paper coding}
    \acro{RIS}{reconfigurable intelligent surface}
    \acro{AWGN}{additive white gaussian noise}
    \acro{MIMO}{multiple-input multiple-output}
    \acro{UL}{uplink}
    \acro{DL}{downlink}
    \acro{OFDM}{orthogonal frequency-division multiplexing}
    \acro{TDD}{time-division duplex}
    \acro{LS}{least squares}
    \acro{MMSE}{minimum mean square error}
    \acro{SINR}{signal to interference plus noise ratio}
    \acro{OBP}{optimal bilinear precoder}
    \acro{LMMSE}{linear minimum mean square error}
    \acro{MRT}{maximum ratio transmitting}
    \acro{M-OBP}{multi-cell optimal bilinear precoder}
    \acro{S-OBP}{single-cell optimal bilinear precoder}
    \acro{SNR}{signal-to-noise ratio}
    \acro{SDR}{Semidefinite Relaxation}
    \acro{SE}{spectral efficiency}
    \acro{GCEs}{Gram channel eigenvalues}
    \acro{BCD}{block coordinate descent}
    \acro{LISA}{linear successive allocation}
\end{acronym}

\title{A Zero-Forcing Approach for the RIS-Aided \\MIMO Broadcast Channel\\}

\author{\IEEEauthorblockN{Dominik Semmler, Michael Joham, and Wolfgang Utschick}
\IEEEauthorblockA{\textit{School of Computation, Information and Technology, Technical University of Munich, 80333 Munich, Germany} \\
email: \{dominik.semmler,joham,utschick\}@tum.de}
}

\maketitle

\begin{abstract}
   We present efficient algorithms for the sum-\ac{SE} maximization of the multi-user \ac{RIS}-aided \ac{MIMO} broadcast channel based on a zero-forcing approach.
   These methods conduct a user allocation for which the computation is independent of the number of elements at the \ac{RIS}, that is usually large.
   Specifically, two algorithms are given that exploit the \ac{LOS} structure between the \ac{BS} and the \ac{RIS}.
   Simulations show superior \ac{SE} performance compared to other linear precoding algorithms but with lower complexity.
\end{abstract}


\begin{figure}[b]
    \onecolumn
    \centering
    \scriptsize{This work has been submitted to the IEEE for possible publication. Copyright may be transferred without notice, after which this version may no longer be accessible.}
    \vspace{-1.3cm}
    \twocolumn
\end{figure}

\section{Introduction}
\label{sec:intro}
\acp{RIS} are under a lot of attention currently as they are possible candidates for supporting future wireless communication systems (e.g., \cite{Power_Min_IRS}). 
Ideally, the \ac{RIS} is an antenna array that is entirely passive and allows the modification of the propagation environment, i.e., shaping the wireless channel of the scenario.
As the array is passive, the idea is to have many antennas at the \ac{RIS} in order to maximize the influence on the channel.

While the \ac{RIS} has the potential to significantly increase the throughput in the system, the complexity of the channel estimation, as well as the complexity of optimizing the phase shift vector at the \ac{RIS}, poses challenges.
In this article, we would like to contribute to the second point and provide an algorithm with reduced complexity to obtain the phase shift vector at the \ac{RIS}.
To this end, we consider perfect \ac{CSI}.
This assumption is often used in the literature, and in this case, it has been shown that including the \ac{RIS} leads to an improvement of the performance in several scenarios.
In \cite{Power_Min_IRS}, for example, it was shown that the power consumption could be clearly reduced, whereas
in \cite{EnergyEff}, the energy efficiency could be improved considerably.

The objective we focus on in this article is the \ac{SE}, which was, for example, also considered in \cite{WSR}, \cite{MaxSumRateJour}, \cite{IRSLISA}.
As linear precoding algorithms are much simpler to realize than, e.g., \ac{DPC} based methods, we restrict ourselves to linear precoding in this article. 
In \cite{WSR}, an alternating optimization based on the WMMSE algorithm was given.
It was, however, mentioned that this algorithm needs a high number of iterations to converge, and, hence, a \ac{BCD} approach was proposed in \cite{WSR}.
To further reduce the complexity of the optimization, the method in \cite{IRSLISA} was developed, which is based on a successive user allocation combined with a zero-forcing approach. 
While this algorithm converges quickly, it is also based on alternating optimization, and the user allocation depends directly on the chosen initial phase shifts.

In this article, we propose a new algorithm that is also based on a successive user allocation as in \cite{IRSLISA}. 
However, instead of relying on alternating optimization,
this algorithm successively adds the users in a greedy manner and, hence, is independent of an initial phase shift.
It is shown in the simulations that this user allocation procedure is more robust and leads to better performance.
A requirement for this particular algorithm, which is not needed for the other algorithms, is that the channel between the \ac{BS} and the \ac{RIS} is dominated by a \ac{LOS} component.
As the \ac{BS} and the \ac{RIS} are, however, intentionally deployed in \ac{LOS} and additionally in considerable height, this assumption is typically fulfilled (see, e.g., \cite{LOSAssump}).

The algorithm exploits this \ac{LOS} structure by performing the complete user allocation within a low-dimensional subspace independent of the number of reflecting elements, which is typically large.
In \cite{Eigenvalues}, it was shown that under the above \ac{LOS} assumption, the eigenvalues of the Gram channel can only be improved to the next larger one of the direct channel with the optimization at the \ac{RIS}.
Only one, the largest eigenvalue of the Gram channel matrix, can be improved without limitations.

We additionally use this result in order to construct an alternative algorithm.
Instead of being based on a complete greedy search, the method exploits that at most one additional user will be served when incorporating the \ac{RIS}.
It should be noted that we consider a scenario where the direct channel is clearly present and non-negligible, and the \ac{RIS} only exists to improve the transmission.
If the channel via the \ac{RIS} were dominant w.r.t. the direct channel, the composite channel matrix would only have a rank of one, and a multi-user scenario is not possible in this case.

\section{System Model}
Throughout this article, we consider a \ac{DL} scenario in which $K$ single-antenna users are served by a single \ac{BS} with $\NB$ antennas.
Furthermore, we assume one \ac{RIS} with $\NRe$ reflecting elements and, hence, the channel from the \ac{BS} to the $k$-th user is denoted as
\begin{equation}
    \bh_k^\Her = \bh^\Her_\tdk + \bh_\trek^\Her \mThet \bH_\ts \; \in \cmplx{1 \times \NB}
\end{equation}
where $\bh_\tdk^\Her \in \cmplx{1 \times \NB}$ is the direct channel from the \ac{BS} to the $k$-th user,  $ \bh_\trek^\Her \in \cmplx{1 \times \NRe}$ is the reflecting channel from the \ac{RIS}
to user $k$, $\mThet = \diag(\bthet) \in \cmplx{\NRe}$ with $\bthet \in \{\bm{z} \in \cmplx{\NRe}: \abs{z_n}=1,\; \forall n\}$ is the phase manipulation at the \ac{RIS}
and 
\begin{equation}
    \bH_\ts = \bm{a}\bm{b}^\Her \in \cmplx{\NRe \times \NB}
\end{equation} is the \ac{LOS} rank-one channel from the \ac{BS} to the \ac{RIS} where w.l.o.g. we assume $\norm{\bm{b}}_2=1$.

We are interested in maximizing the sum \ac{SE} with linear precoding based on a zero-forcing approach.
Zero-forcing precoding is known to be optimal in the high-\ac{SNR} regime.
However, its performance is degraded in the low and medium \ac{SNR} regions.
Moreover, if there are more users to serve than there are \ac{BS} antennas available, it is not possible to construct the zero-forcing filter in its usual form.
In \cite{ZeroForcingUserSelec} and \cite{OriginalLISA}, this was solved by a greedy user allocation.
The zero-forcing filter was then only constructed based on the allocated users.
The user allocation allows the method to perform well in the low and mid \ac{SNR} regions and also provides a solution if there are more users than \ac{BS} antennas.

In comparison to \cite{ZeroForcingUserSelec} and \cite{OriginalLISA}, we consider a scenario including an \ac{RIS} and, hence, the optimization problem we would like to solve is given (see, e.g., \cite{IRSLISA}) by
\begin{equation}
    \label{eq:ActualOptProblem}
    \begin{aligned}
    &\underset{i,\pi,\bm{\Gamma}_i,\bthet}{\max}&& \summe{j=1}{i} \log_2(1+ \lambda_{j,i}\gamma_{j,i})\\
    &\text{s.t.} && \norm{\bm{P}_{\text{eff},i}}_{\text{F}}^2 \le P_{\text{Tx}}
    \end{aligned}
\end{equation}
with the number of allocated users $i$, the user allocation $\pi$ where $\pi(j)$ is the $j$-th allocated user, the composite channel matrix 
\begin{equation}
    \label{eq:CompositeChannelMatrixDefinition}
    \bm{H}_{\text{c},i} =     \underbrace{\begin{bmatrix}
        \bh_{\td,\pi(1)}^\Her\\
        \vdots\\
        \bh_{\td,\pi(i)}^\Her\\
    \end{bmatrix}}_{\bm{H}_{\text{c,d},i}}
    +
    \underbrace{\begin{bmatrix}
        \bh_{\tre,\pi(1)}^\Her\\
        \vdots\\
        \bh_{\tre,\pi(i)}^\Her\\
    \end{bmatrix}}_{\bm{H}_{\text{c,r},i}}
    \bm{\Theta}
    \bm{a}\bm{b}^\Her
\end{equation}
comprising all $i$ allocated users as well as the zero-forcing precoder
\begin{equation}
    \label{eq:EffectivePrecoderDefinition}
\bm{P}_{\text{eff},i} = \bm{H}^+_{\text{c},i} \bm{\Lambda}_i^{\frac{1}{2}} \bm{\Gamma}_i^{\frac{1}{2}}
\end{equation}
where $\bm{\Lambda}_i = \diag(\lambda_{1,i},\lambda_{2,i},\dots,\lambda_{i,i})$ with $\lambda_{j,i}= \frac{1}{\norm{\bm{H}_{\text{c},i}^+\bm{e}_j}^2_2}$ normalizes the column entries of the pseudoinverse to unit-norm
and $\bm{\Gamma}_i = \diag(\gamma_{1,i},\gamma_{2,i},\dots,\gamma_{i,i})$ is the power allocation for the different subchannels, (cf. \cite{IRSLISA}).

Solving \eqref{eq:ActualOptProblem} would require to solve a combinatorial problem.
Additionally, determining the optimal phase vector $\bthet$ and power allocations $\bm{\Gamma}$ for each of the possibilities is clearly infeasible.
Therefore, we will provide a low-complexity algorithm based on a greedy user allocation that showed good performance without the \ac{RIS} in \cite{OriginalLISA}.
The algorithm is based on successively adding users until the inclusion of a user decreases the performance.
In this case, the algorithm is stopped.
An essential aspect of this algorithm is that the user allocation will be performed in a complexity independent of the number of reflecting elements at the \ac{RIS}.
\section{Greedy User Allocation}
As a starting point for the greedy user allocation, we assume that we have allocated $i-1$ out of $K$ possible users and obtained the corresponding \ac{SE} for comparison.
The task is now to find the following user $k=\pi(i)$ and the new power allocations $\bm{\Gamma}_{i}$ jointly with the phase shift vector $\bthet$ that maximize the sum-\ac{SE}.
Mathematically, we would like to solve the optimization problem
\begin{equation}
    \label{eq:GreedyProblem}
    \begin{aligned}
    &\underset{k=\pi(i),\bm{\Gamma}_i,\bthet}{\max}&& \summe{j=1}{i}  \log_2(1+ \lambda_{j,i}\gamma_{j,i})\\
    &\text{s.t.} &&\norm{\bm{P}_{\text{eff},i}}_{\text{F}}^2 \le P_{\text{Tx}}.
    \end{aligned}
\end{equation}
It is important to note that $ \lambda_{j,i}$ as well as the effective precoder $\bm{P}_{\text{eff},i}$  depend on the user allocation $k$ which can be seen from the definitions in \eqref{eq:CompositeChannelMatrixDefinition} and \eqref{eq:EffectivePrecoderDefinition}.
This makes the optimization challenging to solve.
Therefore, we simplify the problem by splitting it into two subtasks.
Firstly, we compute the user allocation based on a lower bound showing promising results in \cite{OriginalLISA} without the \ac{RIS}.
Secondly, we evaluate the new allocation by jointly optimizing the power allocations with the phase shift vector $\bthet$.
\subsection{User Allocation}
\label{sec:UserAlloc}
To find the optimal user in the $i$-th step, we use a tight lower bound (see \cite{OriginalLISA}) on the sum-\ac{SE}.
Incorporating this lower bound, problem \eqref{eq:GreedyProblem} reduces (see \cite{OriginalLISA}) to
\begin{equation}
    \label{eq:LowerBoundGreedy}
    \begin{aligned}
        &\underset{k,\bthet}{\max}&& \left[R_{\text{LB}}  =\log_2\left(1+ \frac{P_{\text{Tx}}}{\norm{\bm{H}^+_{\text{c},i}}_{\text{F}}^2 }\right)\right].\\
        \end{aligned}
\end{equation}
It would also be possible to optimize $\bthet$ w.r.t. the lower bound for every user $k$ and then perform 
the user allocation by evaluating the actual \ac{SE} with an optimization of the power allocations.
However, we have obtained good results by simply using the lower bound for the user allocation as well and, hence, it is used throughout this article.
The problem in \eqref{eq:LowerBoundGreedy} is the same as maximizing the harmonic mean of the composite channel matrix' eigenvalues.
Hence, the optimization will result in improving the condition of the composite channel (see \cite{Eigenvalues}).

Exploiting the \ac{LOS} structure of the \ac{BS}-\ac{RIS} channel we can write the Gram channel matrix (see \cite{Eigenvalues}) as 
\begin{equation}
    \begin{aligned}
        \bm{H}_{\text{c},i}\bm{H}_{\text{c},i}^{\He} &= \bm{C}_i + \bm{D}_i \bbthet \bbthet^\Her \bm{D}_i^\Her,\quad \text{with}\\
    \end{aligned}
\end{equation}
    \begin{align}
        \label{eq:CiDefinition}
        \bbthet &= \begin{bmatrix}
            \bthet^\Her & 1
        \end{bmatrix}^\Her,\quad
    \bm{C}_i  = \bm{H}_{\text{c,d},i}(\eye - \bm{b}\bm{b}^\Her) \bm{H}_{\text{c,d},i}^\Her \; \text{and}\\
    \label{eq:DiDefinition}
     \bD_i &= \begin{bmatrix}
        \bm{H}_{\text{c,r},i}\diag(\bm{a}),\; \bm{H}_{\text{c,d},i}\bm{b}
    \end{bmatrix}.
\end{align}
Using this reformulation and the matrix inversion lemma, we can write the Frobenius norm as 
\begin{equation}
    \label{eq:FrobNormLOS}
    \begin{aligned}
        \norm{\bm{H}^+_{\text{c},i}}_{\text{F}}^2&= \tr\left(\left(\bm{C}_i + \bm{D}_i \bbthet \bbthet^\Her \bm{D}_i^\Her\right)^{\inv}\right)\\
    &=\tr(\bm{C}_i^{\inv}) - \frac{\bbthet^\Her \bm{D}_i^\Her \bm{C}^{-2}_i \bm{D}_i \bbthet}{1 + \bbthet^\Her \bm{D}_i^\Her \bm{C}^{-1}_i \bm{D}_i \bbthet}
    \end{aligned}
\end{equation}
where we assume that the inverse of $\bm{C}_i$ exists.
When $i=\NB$, the inverse does not exist, and this case is considered in subsection \ref{sec:SingularCi}.
With the expression of the Frobenius norm in \eqref{eq:FrobNormLOS}, the maximization of \eqref{eq:LowerBoundGreedy} reduces to
\begin{equation}
    \label{eq:FrobNormOptGreedy}
    \begin{aligned}
    &\underset{k,\bthet}{\min} &&\tr\left(\bm{C}_i^{\inv}\right) - \frac{\bbthet^\Her \bm{D}_i^\Her \bm{C}^{-2}_i \bm{D}_i \bbthet}{1 + \bbthet^\Her \bm{D}_i^\Her \bm{C}^{-1}_i \bm{D}_i \bbthet}.
    \end{aligned}
\end{equation}
As replacing $\bbthet \in \cmplx{\NRe +1 }$ by $\bm{\tilde{\theta}} = \bbthet e^{\mathrm{j}\alpha} \in \cmplx{\NRe +1 }$ does not change the objective for any $\alpha$, we can solve the minimization w.r.t. any unit-modulus constrained vector $\bm{\tilde{\theta}} \in \cmplx{\NRe +1 }$ instead of $\bthet \in \cmplx{\NRe}$.
The actual phase shift vector $\bthet$ can then be recovered by choosing all but the last element of $\bm{\tilde{\theta}}$ and multiplying this vector by $e^{-\mathrm{j}\alpha}$.

To avoid solving \eqref{eq:FrobNormOptGreedy} w.r.t. the unit-modulus constrained $\bm{\tilde{\theta}}$, we relax the vector $\bm{\tilde{\theta}}$ from the set of unit-modulus constrained elements to a spectral norm constraint (i.e., $\norm{\bm{\tilde{\theta}}}_2^2 = \NRe +1$).
By projecting the relaxed solution back to the unit-modulus constraints, this technique can be used to find a possible choice for the phase shift vector $\bthet$ at the \ac{RIS}.
It is, however, important to note that we do not project back on the unit-modulus constraints, and we only use this method to obtain the user allocation and not to get the actual phase shift vector at the \ac{RIS}.
With the spectral norm relaxation, the maximization of \eqref{eq:FrobNormOptGreedy} w.r.t. $\bm{\tilde{\theta}}$ is a conventional generalized eigenvalue problem, and it can be shown that the solution is given by
\begin{equation}
    \label{eq:FrobNormMetric}
    \tr (\bm{C}_i^{\inv}) - \lambda_{\max}\left( \bm{A}_i \right)
\end{equation}
where $\bm{A}_i$ is defined as 
\begin{equation}
    \bm{A}_i = \bm{C}_i^{\inv}-\left(\bm{C}_i+(\NRe+1)\bD_i\bD_i^\Her\right)^{\inv}.
\end{equation}
We can now evaluate \eqref{eq:FrobNormMetric} for the remaining $K-(i-1)$ users and then choose the one with the lowest value.
This process can be performed without computing the optimal value of $\bm{\tilde{\theta}}$.
To avoid the computation of $\bD_i\bD_i^\Her$, which scales in the number of reflecting elements, we apply the QR decomposition
\begin{equation}
    \bD^\Her = \bm{Q} \bm{R}
\end{equation}
once at the beginning of the algorithm.
The matrix $\bD$ is defined as $\bD_i$ in \eqref{eq:DiDefinition} with the difference that all possible users are considered.
We can now construct $\bm{R}_i$ from $\bm{R}$ such that $\bm{R}_i$  only contains the columns corresponding to the allocated users.
It follows that  the computation of $\bD_i\bD_i^\Her $ results in 
\begin{equation}
    \bD_i\bD_i^\Her =  \bm{R}_i^\Her \bm{Q}^\Her \bm{Q} \bm{R}_i  =  \bm{R}_i^\Her \bm{R}_i
\end{equation}
and is therefore independent of the number of \ac{RIS} elements.
The QR decomposition of $ \bD^\Her$ is not independent of the number of reflecting elements but, as mentioned above, needs to be computed only once at the beginning of the algorithm.
The neccessary condition for $\bm{Q}^\Her \bm{Q} = \eye$ is that $ \NRe +1 \ge K$.
Typically, we have many more reflecting elements than users, so this assumption is normally fulfilled.
\subsection{Evaluation}
\label{sec:Evaluation}
Having allocated the user, we need to evaluate the \ac{SE} in order to know when to stop the allocation.
Hence, a joint optimization of the phase vector and the power allocations must be conducted.
Using alternating optimization, the optimization problem for a given phase shift vector reduces to conventional waterfilling.
Hence, we focus on the other subtask in this subsection, i.e., the optimization of the phases at the \ac{RIS} for fixed power allocations $\bm{\Gamma}_{i}$.
Given the fixed power allocations, it is possible to solve the problem (see \cite{IRSLISA})
\begin{equation}
    \label{eq:EvalPhaseOpt}
    \begin{aligned}
    &\underset{\bm{\tilde{\theta}}}{\min}&& \tr\left(\left( \bC_i + \bD_i \bm{\tilde{\theta}} \bm{\tilde{\theta}}^\Her \bD_i^\Her \right)^{\inv}\bm{\tilde{\Gamma}_i}\right)\\
    \end{aligned}
\end{equation}
where $\bm{\tilde{\Gamma}}_i = \bm{\bar{\Lambda}}_i \bm{{\Gamma}}_i$ and $\bm{\bar{\Lambda}}_i$ are the channel gains given by the phase shift vector of the last iteration.
In \eqref{eq:EvalPhaseOpt} a phase vector is found which achieves the same sum-\ac{SE} for a lower transmit power.
The element-wise algorithm given in \cite{IRSLISA} takes into account the unit-modulus constraints and shows a fast convergence behavior for optimizing \eqref{eq:EvalPhaseOpt}.
However, applying the alternating optimization with the element-wise update of $\NRe$ elements in each iteration, together with optimizing the power allocations, is computationally expensive when considering a higher number of users
as the whole optimization has to be performed once a new user is added.

By applying again the spectral norm relaxation as in the last subsection, it is possible to perform the optimization within a low-dimensional subspace which is independent of $\NRe$.
In particular, we change the optimization variable to 
\begin{equation}
    \bm{u} = \bm{Q}^\Her \bm{\tilde{\theta}} \in \cmplx{K}.
\end{equation}
This is possible, as when given the optimal $\bm{u}^{\star}$, we can directly obtain the optimal $\bm{\tilde{\theta}}^{\star}$ under the spectral norm relaxation by $\bm{\tilde{\theta}}^{\star} = \bQ \bm{u}^{\star}$ where the requirement is again $K \le \NRe + 1$.
The resulting optimization problem can be expressed as
\begin{equation}
    \begin{aligned}
    &\underset{\bm{{u}}}{\min}&& \tr\left(\left( \bC_i + \bm{R}_i^\Her \bm{u} \bm{u}^\Her \bm{R}_i\right)^{\inv} \bm{\tilde{\Gamma}}_i\right).\\
    \end{aligned}
\end{equation}
Under the spectral norm relaxation, the above problem is similar to \eqref{eq:FrobNormOptGreedy}, a generalized eigenvalue problem 
and it can be shown that the optimal solution is given by
\begin{equation}
    \tr (\bm{C}_i^{\inv}\bm{\tilde{\Gamma}_i}) - \lambda_{\max}( \bm{A}_i \bm{\tilde{\Gamma}}_i).
\end{equation}
The updated channel gains can now be expressed as
\begin{equation}
    \label{eq:EvalUpdatedPowerGains}
    \lambda_{j,i} = \frac{1}{\bm{e}_j^{\T} (\bm{C}_i^{\inv}-\bm{v}\bm{v}^\Her\bm{A}_i )\bm{e}_j}
\end{equation}
where $\bm{v}$ is the principal eigenvector of $\bm{A}_i\bm{\tilde{\Gamma}}_i$.
The vector $\bm{u}$ can always be recovered (e.g., after the convergence of the algorithm to obtain an initial solution for $\bthet$) by
 $\bm{u} = \frac{ \bm{R}_i \left(\bm{C}_i+(\NRe+1) \bm{R}_i^\Her \bm{R}_i\right)^{\inv}\bm{\tilde{\Gamma}_i}\bm{v}}{\norm{ \bm{R}_i \left(\bm{C}_i+(\NRe+1) \bm{R}_i^\Her \bm{R}_i\right)^{\inv}\bm{\tilde{\Gamma}_i}\bm{v}}_2} \sqrt{\NRe + 1}$.
It is now possible to compute the new power allocations by performing waterfilling with the channel gains $ \lambda_{j,i}$ given in \eqref{eq:EvalUpdatedPowerGains}.
During the waterfilling subtask, some users might get no power allocated. 
We deallocate all the users during this optimization that receive no power.

As an initial point for the alternating optimization, we take the scaled identity for $\bm{\tilde{\Gamma}}_i$, which is the solution of the lower bound used in Section \ref{sec:UserAlloc} for the user allocation.

It is now possible to give the sum-\ac{SE} for the spectral norm relaxation as
\begin{equation}
    \label{eq:EvalSumSE}
    \mathrm{SE} = \summe{j=1}{i} \log_2\left( 1 + \lambda_{j,i}\gamma_{j,i}\right)
\end{equation}
which is used as a stopping criterion, i.e., if the \ac{SE} does not improve, the user allocation is stopped.
\subsection{Singular $\bC_i$, i.e. $i = \NB$}
\label{sec:SingularCi}
So far, we only considered the case in which $\bm{C}_i$ is invertible.
However, when $i=\NB$ the multiplication with the orthogonal projector in \eqref{eq:CiDefinition}
results in the matrix $\bm{C}_i$ having one eigenvalue equal to zero and the expression
\begin{equation}
    \label{eq:RankDefCiFrobNormPowerAlloc}
    \tr\left(\left(\bC_i + \bm{D}_i\bm{\tilde{\theta}} \bm{\tilde{\theta}}^\Her \bm{D}_i^\Her\right)^{\inv}\bm{\tilde{\Gamma}}_i\right),
\end{equation}
which is needed in subsection \ref{sec:Evaluation} and with $\bm{\tilde{\Gamma}}_i = \eye$ in subsection \ref{sec:UserAlloc}, cannot be calculated in the usual way.
However, by exploiting the result of \cite{pseudoinv},  \eqref{eq:RankDefCiFrobNormPowerAlloc} can be reformulated as 
\begin{equation}
    \begin{aligned}
         \tr(  \bm{\tilde{C}}_i^+)  +   \frac{1+ \bm{\tilde{\theta}}^{\Her}\bm{\tilde{D}}_i^{\Her} \bm{\tilde{C}}_i^+ \bm{\tilde{D}}_i\bm{\tilde{\theta}} }{\norm{\bm{w}^{\Her}\bm{\tilde{D}}_i\bm{\tilde{\theta}}}_2^2} \\
    \end{aligned}  
\end{equation}
where $\bm{\tilde{C}}_i = \bm{\tilde{\Gamma}}_i^{-\frac{1}{2}}\bC\bm{\tilde{\Gamma}}_i^{-\frac{1}{2}}$, $\bm{\tilde{D}}_i = \bm{\tilde{\Gamma}}_i^{-\frac{1}{2}}\bm{{D}}_i$ and $\bm{w}$ is the eigenvector
corrsponding to the zero-eigenvalue of $\bm{\tilde{C}}_i$.
It can be directly inferred that $\bm{\tilde{\theta}} = \frac{\left(\eye \frac{1}{\NRe +1} + \bm{\tilde{D}}_i^{\Her} \bm{\tilde{C}}_i^+ \bm{\tilde{D}}_i\right)^{\inv}\bm{\tilde{D}}^\Her_i\bm{w}}{\norm{\left(\eye \frac{1}{\NRe +1} + \bm{\tilde{D}}_i^{\Her} \bm{\tilde{C}}_i^+ \bm{\tilde{D}}_i\right)^{\inv}\bm{\tilde{D}}^\Her_i\bm{w}}_2}\sqrt{\NRe + 1}$ is the optimal solution under the spectral norm constraint.

Having reformulated the expression \eqref{eq:RankDefCiFrobNormPowerAlloc} in a similar form to \eqref{eq:FrobNormOptGreedy},
it is possible to obtain slightly adjusted versions (also within the low-dimensional subspace) of the solutions in subsection \ref{sec:UserAlloc} and \ref{sec:Evaluation} where $\bC_i$ has full rank.
\section{User-Allocation Algorithm}

We now describe two algorithms where one is based on a complete greedy search, whereas the other one exploits the user selection of the direct channel by using the eigenvalue result of \cite{Eigenvalues}.
\subsection{Complete Greedy Search}
\label{sec:CompleteGreedy}
The first algorithm, we examine, is based on a complete greedy search.
This means that we start with zero users, i.e., $i=0$, and then successively add users until the sum-\ac{SE} according to \eqref{eq:EvalSumSE} does not increase anymore.
After the user-allocation has stopped, the number of users (in the following denoted as $d$), the user allocation $\pi$, as well as the power allocations $\bm{\Gamma}_d$, have been obtained.
We are now able to find the optimal vector $\bthet$ with updated power allocations $\bm{\Gamma}_d$.

In particular, we solve the same optimization problem as in subsection \ref{sec:Evaluation} and perform alternating optimization between obtaining the updated power allocations by waterfilling for a fixed $\bthet$
and computing the new phase shift vector $\bthet$ for fixed power allocations by 
\begin{equation}
    \label{eq:CompleteGreedyPhaseOpt}
    \begin{aligned}
    &\underset{\bthet}{\min}&& \tr\left(\left( \bC_d + \bD_d \bbthet \bbthet^\Her \bD_d^\Her \right)^{\inv}\bm{\tilde{\Gamma}}_d\right)\\
    \end{aligned}
\end{equation}
where the channel gains based on the phase shift vector of the last iteration are again denoted by $\bm{\bar{\Lambda}}_d$ and $\bm{\tilde{\Gamma}}_d = \bm{\bar{\Lambda}}_d \bm{{\Gamma}}_d$.
In subsection \ref{sec:Evaluation}, a solution on the spectral norm relaxation within a low-dimensional subspace was obtained.
However, this solution was only used in order to evaluate the user allocation for the stopping criteria, whereas here, we would like to obtain an accurate solution because the actual phase vector $\bthet$ for the \ac{RIS} is designed.
Hence, this time, we perform the calculation within the original dimension $\NRe$ and take into account the unit-modulus constraints by using the element-wise algorithm given in \cite{IRSLISA}.
The updated channel gains are then computed by 
\begin{equation}
    \begin{aligned}
    \lambda_{j,d} = \frac{1}{\bm{e}_j^{\T}\bC_d^{-1}\bm{e}_j - \frac{\abs{\bbthet^\Her \bm{D}_d^\Her \bm{C}^{-1}_d \bm{e}_j}^2 }{1 + \bbthet^\Her \bm{D}_d^\Her \bm{C}^{-1}_d \bm{D}_d \bbthet}}
    \end{aligned}
\end{equation}
and waterfilling can be applied to find the updated power allocations.
As an initial guess, we use $\bm{\tilde{\Gamma}}_d$ from the spectral norm relaxation in subsection \ref{sec:Evaluation} and the corresponding projected phase shift solution
\begin{equation}
    \bm{\tilde{\theta}} = \exp\left(\mathrm{j}\arg\left(\bm{Q}\bm{u}\right)\right)
\end{equation}
from which $\bthet$ can be inferred.
In this article, we update all phase shifts in the optimization problem \eqref{eq:CompleteGreedyPhaseOpt} only once according to the algorithm from \cite{IRSLISA} and then update the power allocations by waterfilling.
Also, a higher number of iterations for the phase shift update could be considered.

While the above algorithm converges fast (see \cite{IRSLISA}) it has a linear scaling in $\NRe$ in comparison to the relaxed solution in subsection \ref{sec:Evaluation}.
This optimization, however, has to be calculated only once at the end of the algorithm.

After the phase shift vector $\bthet$ is obtained, we can apply any conventional algorithm to get the precoders at the \ac{BS}.
We opt for the conventional \ac{LISA} algorithm, after which the optimization is finished.

\subsection{AddOne-RIS-LISA}
While the user allocation is independent of $\NRe$, it is still computationally expensive to perform a complete greedy search as suggested above.
Hence, in this subsection, we provide an algorithm with reduced complexity by using the eigenvalue limitation result of \cite{Eigenvalues} for the \ac{LOS}-dominated \ac{BS}-\ac{RIS} channel.
Here, it was stated that the eigenvalues of the channel $ \lambda_n$ (we assume a decreasing order of the eigenvalues throughout this article) are bounded (see \cite{Eigenvalues}) by 
\begin{equation}\label{eq:RankOneLimit}
    \begin{aligned}
        \lambda_n & \le \lambda^\td_{n-1},  \; n=2,\dots,i\;
            & \text{and} \; \lambda_1 \; \text{being unbounded}
    \end{aligned}
\end{equation}
where $\lambda^\td_{n}$ are the eigenvalues of the direct channel and $\lambda_{n}$ the eigenvalues of the channel including the \ac{RIS}. 
This means that the \ac{RIS} can only increase the eigenvalues of the channel to the next larger one of the direct channel and additionally, it can only increase one eigenvalue without limitations.
Applying the above restrictions on the composite channel matrix $\bH_{\text{c},i}$ in the $i$-th step, we obtain (for all user allocations)
\begin{equation}\label{eq:RankOneLimit}
    \begin{aligned}
        \lambda^{}_n(\bH_{\text{c},i}) & \le \lambda^{\td}_{n-1}(\bH_{\text{c,d},i}),  \; n=2,\dots,i\;\\
            & \text{and} \;  \lambda^{}_1(\bH_{\text{c},i}^{}) \; \text{being unbounded}.
    \end{aligned}
\end{equation}

In the following, we assume that we have obtained the user allocation based on only the direct channel (e.g., with \ac{LISA} \cite{OriginalLISA}),
where we denote the number of allocated users as $i_{\text{D}}$. 
Two more users are now allocated, and the behavior of the eigenvalues is analyzed.
After the inclusion of the two additional users, we know, with the result of \cite{Eigenvalues}, that the eigenvalue bounds
\begin{equation}\label{eq:RankOneLimit}
    \begin{aligned}
        \lambda^{}_{i_{\text{D}}+1}(\bH_{\text{c},i_{\text{D}}+2}) & \le \lambda^{\td}_{i_{\text{D}}}(\bH_{\text{c,d},i_{\text{D}} +2}), \\
        \lambda^{}_{i_{\text{D}}+2}(\bH_{\text{c},i_{\text{D}}+2}) & \le \lambda^{\td}_{i_{\text{D}} +1}(\bH_{\text{c,d},i_{\text{D}} + 2})\\
    \end{aligned}
\end{equation}
hold for any user allocation.
We can see that $\lambda^{}_{i_{\text{D}}+2}(\bH_{\text{c},i_{\text{D}}+2})$ can not exceed $\lambda^{\td}_{i_{\text{D}} +1}(\bH_{\text{c,d},i_{\text{D}} + 2})$
which was not large enough to be allocated during the user allocation of the direct channel.

From the discussion above, we conclude that the \ac{RIS} will, at maximum, lead to the inclusion of one additional user.
Hence, instead of performing a complete greedy search, we only check the allocations with $i_{\text{D}}$ and $i_{\text{D}} + 1$ users, i.e., where no or one user is added.

At first, we compute the user allocation for $i_{\text{D}}$ users with the inclusion of the \ac{RIS}.
Knowing that the \ac{RIS} leads at maximum to the inclusion of one additional user, we deallocate the last (worst) user of the direct channel allocation and then add a user by performing the joint optimization with the \ac{RIS} described in subsection \ref{sec:UserAlloc}.
The joint optimization is the reason for first deleting and then adding a user to the allocation.

Next, we compute the user allocation for $i_{\text{D}} + 1$ users.
For this, we take the allocation of the direct channel as a basis, even though the new allocation obtained above by first removing and then adding a user could be used as well.
The performance of the two approaches was very similar, and throughout this article, we focus only on the case where the direct channel allocation is used.
Hence, an additional user is allocated to the allocation of the direct channel by performing the joint optimization with the \ac{RIS} of subsection \ref{sec:UserAlloc}.

This algorithm only has to perform the user allocation twice instead of conducting a complete greedy search.
We now only have to evaluate the allocations for $i_{\text{D}}$ and $i_{\text{D}}+1$ users and then take the better one.
Here, we take again the low-complexity evaluation of subsection \ref{sec:Evaluation} where the evaluation is performed within the low-dimensional subspace.
However, as only two cases have to be checked, one could further improve the algorithm by evaluating the two allocations in the original dimension $\NRe$ as it was described in subsection \ref{sec:CompleteGreedy} by alternating optimization with waterfilling and solving \eqref{eq:CompleteGreedyPhaseOpt} with the algorithm in \cite{IRSLISA}.
We only experienced a slight performance increase, and throughout this article, we will conduct the evaluation according to subsection \ref{sec:Evaluation} within the low-dimensional subspace.
Additionally, this allows us to better compare the method to the complete greedy search as now both algorithms use the same evaluation method and only differ in their user selection.

During the evaluation of the allocations, users might also be deallocated here when they receive no power during the waterfilling subtask.
This is less likely when the direct channel is dominant and the \ac{RIS} improves the smallest eigenvalue.
We are also mainly considering this scenario in this article. With a dominant direct channel, multiple users can be served even though the channel via the \ac{RIS} has a rank of one.

\paragraph*{Complexity} As the user allocation is independent of $\NRe$ and only two allocations have to be evaluated, this algorithm has a very low complexity.
The joint optimization of $\bthet$ and $\bm{\Gamma}$ at the end of the algorithm converges extremely quickly once the correct user allocation has been found.
\section{Simulations}
For the simulations we consider a \ac{BS} with $\NB = 8$ antennas at $(0\,\text{m}, 0\,\text{m}, 10\,\text{m})$ together with an $\ac{RIS}$ at $(100\,\text{m}, 0\,\text{m}, 10\,\text{m})$ which has $\NRe$ (values will be given in the plots) elements.
The users are assumed to be on a height of $1.5 \, \text{m}$ and uniformly distributed in a circle with radius $5\,\text{m}$ centered at $(95\,\text{m}, 10\,\text{m}, 1.5\,\text{m})$ where all have a single antenna at their disposal.
We use the logarithmic path loss model $ L_{\text{dB}} = \alpha + \beta 10\log_{10}(\frac{d}{\mathrm{m}})$ for all of the channels where $d$ is the distance in meter.
For the direct channel, the channel between the \ac{RIS} and the users, as well as the channel between the \ac{BS} and the \ac{RIS}, we set $\alpha_\td = \alpha_\tre = \alpha_\ts=30\,\mathrm{dB}$ for the reference distance of $\text{1m}$ whereas 
$\beta_\tdk=3.7$, $\beta_\trek=3.2$ and $\beta_\ts=2.2$ for the distance dependent path loss.
In all simulations, we additionally assume a noise variance of $\sigma^2=-100\,\text{dBm}$ at the receive antennas.
The direct channel, as well as the channel from the \ac{RIS} to the users, is modeled as uncorrelated Rayleigh fading, 
whereas the channel between the \ac{BS} and the \ac{RIS} is given by a pure \ac{LOS} channel.
For this, we assume a $\frac{\lambda}{2}$ spaced \ac{ULA} at the \ac{BS} and the \ac{RIS} where the angle of arrival/departure is given by $0$ degrees.

The number of users is switched between $K = 4 < \NB$ and $K= 12 > \NB$, and the actual number is given in the plots.
In order to have a more challenging setup, we assume that the direct channel for half of the users ($2$ or $6$ users, respectively) has an additional path loss of $20\text{dB}$.

Furthermore, we consider the following algorithms. 
We use the \ac{LISA} (see \cite{OriginalLISA}) algorithm for evaluating the performance of the direct channel and the random phase shifts.
Additionally, we use the \ac{DPC}-AO from \cite{MaxSumRateJour} for evaluating the upper bound.
In particular, the algorithm is started $10$ times with different random phase shifts as initial guesses, and then the best is taken.
Moreover, we consider the \ac{RIS}-\ac{LISA} algorithm of \cite{IRSLISA}, which is also based on \ac{LISA} but with alternating optimization.
The initial phase shift is given by the same random phase shifts as the random phase shifts, which are plotted in the figures.
Furthermore, the \ac{BCD} of \cite{WSR} with the same initial phase shift is given where we use as a stopping criteria a maximum iteration number of $2000$.
From this paper, we evaluate the Greedy-RIS-LISA as well as the AddOne-RIS-LISA.

\subsection{Less Users than BS antennas}
\begin{figure}[h!]
    \flushleft
    \hspace*{18pt}
    \vspace*{-20pt}
     \includegraphics[scale=1]{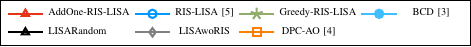}
    \flushleft

    \subfigure[$\NRe = 128$]{
        \hspace*{-8pt}
        \includegraphics[scale=1]{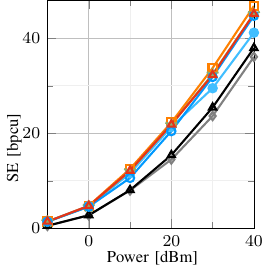}

    \label{fig:MoreAntennasPower}

    }\hspace*{-10pt}\subfigure[$P_{\text{Tx}}=20\text{dBm}$]
    {

        \includegraphics[scale=1]{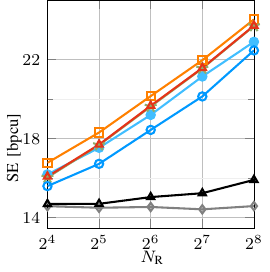}
        \label{fig:MoreAntennasElements}

    \vspace{-5pt}
    }
    \caption{ \ac{SE} evaluation for $K=4$ users.}
    \label{fig:MoreAntennas}

\end{figure}

In Figure \ref{fig:MoreAntennas}, we can see the performance for $K=4$ users where half of them have an additional path loss of $20\text{dB}$ for the direct channel.
All algorithms show a clear performance gain over the random phase shifts and just the direct channel in Figure \ref{fig:MoreAntennasPower} for the complete power range.
Even though restricted by the eigenvalue limitations \cite{Eigenvalues}, such a significant gain can be expected for a low number of users.
With only two users having a good direct channel, improving just one of the remaining two has a strong impact on the sum-\ac{SE}.

Additionally, we can see that the \ac{BCD} algorithm has a good performance in the low and mid power range but experiences a degradation for higher power values.
The reason for this is that the algorithm needs a lot of iterations for higher power values and, hence, the maximum iteration number limits its performance.
When considering the algorithms based on \ac{LISA}, we can see that the Greedy-RIS-LISA, as well as the AddOne-RIS-LISA perform similarly, whereas the algorithm of \cite{IRSLISA} 
has a slight degradation for mid-power values.
This is further shown in Figure \ref{fig:MoreAntennasElements} for a varying number of reflecting elements.
Here, the Greedy-RIS-LISA and AddOne-RIS-LISA algorithms are performing best, followed by the \ac{BCD} algorithm,
whereas the algorithm presented in \cite{IRSLISA} has some gap to the other algorithms.

\subsection{More Users than \ac{BS} Antennas}
\begin{figure}[h!]
    \flushleft
    \hspace*{18pt}
    \vspace*{-20pt}
    \includegraphics[scale=1]{Figures/Legend.pdf}
    \flushleft
\subfigure[$\NRe = 128$]{
        \hspace*{-8pt}
        \includegraphics[scale=1]{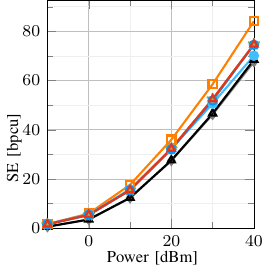}
         \label{fig:MoreUsersPower}
    
    }\hspace*{-10pt}\subfigure[$P_{\text{Tx}}=20\text{dBm}$]{    
        \includegraphics[scale=1]{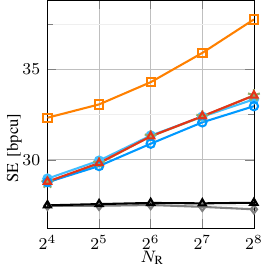}

            \label{fig:MoreUsersElements}

    }
    \caption{\ac{SE} evaluation for $K=12$ users.}
    \label{fig:MoreUsers}
\vspace{-5pt}
\end{figure}

We now analyze the performance of the algorithms in Figure \ref{fig:MoreUsers} for $K=12$ where again half of the users have an additional $20\text{dB}$ path loss for the direct channel.
In Figure \ref{fig:MoreUsersPower} a similar situation to the 4-user scenario above arises.
However, we can see a less prominent gain in comparison to the direct channel for all of the algorithms.
The reason is that now $6$ of the $12$ users have a good channel condition, and relatively improving one of the remaining users has a smaller impact than in the scenario above.
No fundamental differences can be seen for the algorithms compared to the $4$ user scenario, except that the gaps are generally closer. 
We also see a similar situation for the number of reflecting elements in Figure \ref{fig:MoreUsersElements}.
The \ac{BCD} is best for fewer elements and performs equal to the Greedy-RIS-LISA and AddOne-RIS-LISA algorithms for a higher number of elements.
\ac{RIS}-\ac{LISA} from \cite{IRSLISA} is only marginally worse, but a gap can be seen over the whole range.

\section{Conclusion}
We have seen that the presented algorithms show a good performance and are independent of any initial guess.
The algorithms work for the \ac{LOS} assumption between the \ac{BS} and the \ac{RIS}, which is likely to be fulfilled.
While the eigenvalue restrictions given in \cite{Eigenvalues} hold for this assumption, we have seen that, in particular, for a low number of users, the \ac{RIS} can still have a significant impact on the scenario.
Nevertheless, the algorithms in this article will be extended to a multi-\ac{RIS} scenario such that the influence of the \acp{RIS} is also maintained when a higher number of users is considered.

\bibliographystyle{IEEEtran}
\bibliography{refs}
\end{document}